\documentstyle[prl,aps,amsfonts,amssymb,epsfig]{revtex}
\begin{document}
\draft 
\twocolumn[\hsize\textwidth\columnwidth\hsize\csname 
@twocolumnfalse\endcsname
\title{The symmetry problem in $\alpha^{\prime}$-NaV$_2$O$_5$ }
\author{A. Damascelli$^a$, D. van der Marel$^a$, J. Jegoudez$^b$, G. Dhalenne$^b$, 
and A. Revcolevschi$^b$}
\address{$^a$Solid State Physics Laboratory, University
 of Groningen, Nijenborgh 4, 9747 AG Groningen, The Netherlands}
\address{$^b$Laboratoire de Chimie des Solides, Universit$\acute{e}$ de 
Paris-sud, 
B$\hat{a}$timent 414, F-91405 Orsay, France}
\date{29 May 1998}
\maketitle
\begin{abstract}
We discuss the symmetry of $\alpha^{\prime}$-NaV$_2$O$_5$ in the high temperature phase 
on the basis of optical conductivity data. Conclusive information cannot be obtained by studying 
the optically allowed lattice 
vibrations. However,  intensity and  polarization of the  electronic 
excitations give a direct indication for a broken-parity electronic ground-state. This 
is responsible for the detection  of {\em charged bi-magnons} in the optical spectrum.
\end{abstract}
\pacs{Keywords:  $\alpha^{\prime}$-NaV$_2$O$_5$; CuGeO$_3$; 
spin-Peierls transition; charged magnons}
\vskip2pc]
\narrowtext
The symmetry of $\alpha^{\prime}$-NaV$_2$O$_5$ in the high temperature phase is the subject of 
intense discussion\cite{carpy,meetsma,PRL} because of its consequences on the 
interpretation of the phase transition at 34 K,  ascribed to a spin-Peierls 
transition involving  one-dimensional (1D) chains of V$^{4+}$\cite{isobe}. This 
interpretation could be consistent with the non-centrosymmetric 
space group {\em P2$_1$mn} originally proposed by Carpy and Galy\cite{carpy}. 
The structure of the compound consists of two-leg ladders running along the {\em b} axis, 
with the rungs 
oriented along the {\em a} axis and defined by two V ions, one on each leg of the ladder. In 
the space-group {\em P2$_1$mn}, the V {\em d}-electrons are distributed in such a way that 
the left and right legs of a ladder are formed by V$^{4+}$ (S=1/2) and V$^{5+}$ (S=0) ions, 
respectively. However, 
this structural analysis has  recently been questioned and the centrosymmetric space group 
{\em Pmmn}  has been proposed\cite{meetsma}: the V ions would have an 
average charge of +4.5 and it would not be possible to identify well distinct 1D magnetic 
chains. Therefore, the interpretation of the phase transition would not be straightforward.

A possible way to assess the symmetry issue is to investigate the Raman (R) and infrared
(IR) phonon spectra comparing the number of experimentally observed modes with the number
expected for the two space groups on the basis of a group-theory analysis. As a 
result of such a calculation we obtain for the irreducible representation of the optical vibrations
for {\em P2$_1$mn}:

\begin{eqnarray*}
\rm{\Gamma} &=& 15\rm{A}_{1}(\rm{aa,bb,cc};{\rm{E}}\|\rm{a})+8\rm{A}_{2}(\rm{bc})
    +7\rm{B}_{1}(\rm{ab};{\rm{E}}\|\rm{b}) \\
&+& 15\rm{B}_{2}(\rm{ac};{\rm{E}}\|\rm{c})~,
\end{eqnarray*} 

\noindent
corresponding to 45 R (A$_{1}$,A$_{2}$,B$_{1}$,B$_{2}$) and 37 IR 
(A$_{1}$,B$_{1}$,B$_{2}$) active modes, and for {\em Pmmn}:

\begin{eqnarray*}
 \rm{\Gamma}^{\prime} &=& 8\rm{A}_{\rm{g}}(\rm{aa,bb,cc})+3\rm{B}_{\rm{1g}}(\rm{ab})+8\rm{B}
_{\rm{2g}}(\rm{ac})+5\rm{B}_{\rm{3g}}(\rm{bc})   \\
&+& 7\rm{B}_{\rm{1u}}({\rm{E}}\|\rm{c})+4\rm{B}_{\rm{2u}}
({\rm{E}}\|\rm{b})+
7\rm{B}_{\rm{3u}}({\rm{E}}\|\rm{a})~,                                  
\end{eqnarray*} 

\noindent
corresponding to 24 R (A$_{\rm{g}}$,B$_{\rm{1g}}$,B$_{\rm{2g}}$,B$_{\rm{3g}}$) and 18 IR
(B$_{\rm{1u}}$,B$_{\rm{2u}}$,B$_{\rm{3u}}$) active modes. However, the number of observed 
phonons\cite{PRL,fischer}  is smaller than that  calculated for both space groups, meaning that some 
of the optical vibrations in $\alpha^{\prime}$-NaV$_2$O$_5$ have a very small oscillator
\begin{figure}[h]
\centerline{\epsfig{figure=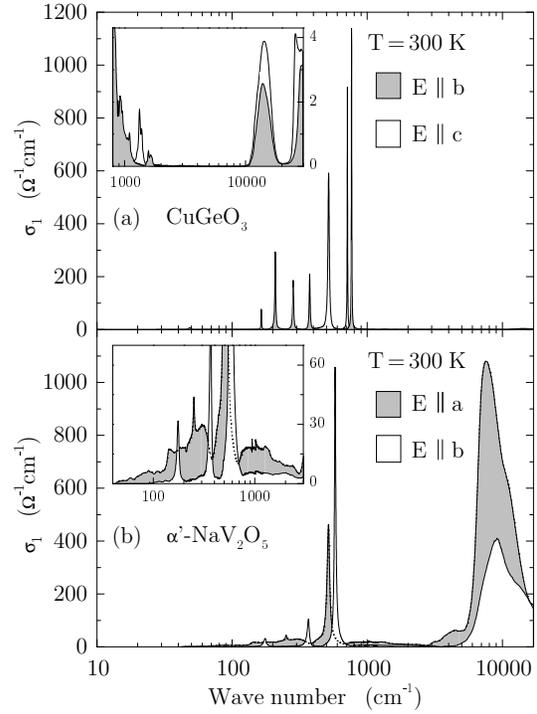,width=7cm,clip=}}
 \caption{Panel (a): optical conductivity of CuGeO$_3$ at 300 K for 
 $\vec{E}\!\parallel\!\vec{b}$ (normal to the chains) and $\vec{E}\!\parallel\!\vec{c}$ 
 (along the chains).  Inset: enlarged  view of $\sigma_{1}(\omega)$ from 800 to 
 $30\,000$ cm$^{-1}$. Panel (b): optical 
 conductivity of $\alpha^{\prime}$-NaV$_2$O$_5$ at 300 K for 
 $\vec{E}\!\parallel\!\vec{a}$ (normal to the ladders) and $\vec{E}\!\parallel\!\vec{b}$ 
 (along the ladders). Inset: enlarged 
 view of $\sigma_{1}(\omega)$ from 40 to 3000 cm$^{-1}$. 
}
\label{fig1}
\end{figure}
\noindent 
strength. Therefore, an unknown number of modes has escaped detection in the 
experiments\cite{PRL,fischer} and none of the two space groups can be ruled out. 

A deeper insight of the symmetry of $\alpha^{\prime}$-NaV$_2$O$_5$ is obtained from  
optical conductivity data [Fig.~1(b)], in particular when compared to  the
result relative to CuGeO$_3$ [Fig.~1(a)].  On the latter compound we observed
sharp phonon lines below 1000 cm$^{-1}$\cite{sp}, multiphonon absorptions at $\sim$1500 cm$^{-1}$,
very weak [note the  low values of $\sigma_1(\omega)$ in the inset of Fig.~1(a)] 
phonon-assisted Cu {\em d-d} transitions at $\sim$$14\,000$ cm$^{-1}$\cite{bassi}, 
and the onset of the Cu-O charge-transfer (CT) excitations at $\sim$$26\,000$ cm$^{-1}$.  On 
$\alpha^{\prime}$-NaV$_2$O$_5$, besides the phonon lines in the far-infrared region, we detected 
features that are completely absent in CuGeO$_3$: a strong absorption peak at $\sim$8000 
cm$^{-1}$ and a low-frequency continuum for $\vec{E}\!\parallel\!\vec{a}$ [inset of Fig.~1(b)]. 
On the basis of intensity considerations, the peak at $\sim$8000 cm$^{-1}$ has 
to be ascribed to an optically allowed excitation and  cannot be interpreted as a V {\em d-d} 
exciton, a transition that is in principle optically forbidden 
and only weakly allowed in the presence of a strong crystal field or electron-phonon coupling. 
The continuum along the {\em a} axis, as the frequency range coincides with the 
low-energy scale spin excitations, has to be due to double spin-flip excitations. Another reason 
for this assignment is the opening,  for T$<$34 K\cite{PRL}, of a gap in the optical 
conductivity of 17$\pm$3 meV ({\em i.e.}, approximately twice the spin gap value\cite{fujii}).
Moreover,  intensity and  polarization  ($\vec{E}\!\parallel\!\vec{a}$)  of the 
continuum cannot be understood assuming the complete equivalence of the V sites required by 
the space group {\em Pmmn}. In fact, direct and phonon-assisted spin excitations, if 
characterized by a finite intensity, would be optically active along the ladders 
($\vec{E}\!\parallel\!\vec{b}$).

A way to understand qualitatively and quantitatively the electronic 
excitations in $\alpha^{\prime}$-NaV$_2$O$_5$ is to assume a broken left-right symmetry of the 
ladders, {\em i.e.}, a difference $\Delta$ between the on-site 
energies of the two V sites on the same rung. Each  rung can  be modeled 
as a polar diatomic molecule with, {\em e.g.}, left bonding ($L_B$) and right anti-bonding 
($R_{AB}$) lob-sided wave functions. An electron can then be optically excited from the $L_B$ 
to the $R_{AB}$ orbital and it is this on-rung CT between the two V sites that we identify with 
the strong absorption 
peak at  $\sim$8000 cm$^{-1}$. From the integrated intensity of the CT absorption 
we can calculate $\Delta \!\approx\! 0.8$ eV and the on-rung hopping parameter 
$|t_{\bot}|\!\approx 0.3$ eV, and  show that the valence of the two V ions on a rung is 4.1 and 
4.9, respectively\cite{PRL}. We now consider a small segment of the ladder with three rungs,
and one electron per rung in the $L_B$ orbital. If the spins are anti-parallel, the middle 
electron experiences some virtual hopping to the two neighboring rungs and, for 
$U\!\rightarrow\!\infty$, it will eventually reside
in a $R_{AB}$ orbital. If the spins are parallel ({\em i.e.}, if the middle spin has been 
flipped over), no virtual hopping is possible because of the Pauli principle. As 
a result, there is a net dipole displacement perpendicular to the legs between the two 
different configurations: spin-flip  excitations carry a finite dipole moment which is 
responsible for the detection  of {\em charged bi-magnons} in the optical spectrum for 
$\vec{E}\!\parallel\!\vec{a}$, {\em i.e.}, 
a  {\em direct two-magnon optical absorption}. It can be shown that the effective charge for this 
process is $q_m\!=\!q_e \frac{3J_{\parallel}\Delta}{\Delta^2+4t_{\perp}^2}$ , 
where $q_e$ is the electron charge and 
$J_{\parallel}$ 
 is the exchange coupling constant between two spins on neighboring rungs\cite{PRL}. 
Therefore, for a symmetrical ladder, where $\Delta$=0, $q_m$=0 and the charged magnon effect 
disappears. On the other hand, for an 
asymmetrically charged ladder $q_m$$\neq$0. It is  clear from the   
argument above that  the symmetry has to  be broken only locally  to obtain optical activity of 
two-magnon excitations: it is sufficient to have short chain segments of at least three spins, 
which are randomly distributed between left and right legs of the ladders. 

In conclusion, by analyzing the optical conductivity, we showed that  intensity and  
polarization of the electronic excitations give a direct evidence for a 
broken-parity electronic ground-state of $\alpha^{\prime}$-NaV$_2$O$_5$ in the high 
temperature phase. The broken symmetry is responsible for the detection of charged bi-magnons 
in the optical spectrum.

\end{document}